*Research Article*

# A New Computer-Aided Diagnosis System with Modified Genetic Feature Selection for BI-RADS Classification of Breast Masses in Mammograms

**Said Boumaraf**,[1] **Xiabi Liu**,[1] **Chokri Ferkous**,[2] and **Xiaohong Ma**[3]

[1]*Beijing Lab of Intelligent Information Technology, School of Computer Science and Technology, Beijing Institute of Technology, Beijing, China*
[2]*Laboratoire des Sciences et Technologies de l'Information et de la Communication LabSTIC, Université 8 Mai 1945 Guelma, BP 401, Guelma 24000, Algeria*
[3]*National Cancer Center/National Clinical Research Center for Cancer/Cancer Hospital, Chinese Academy of Medical Sciences and Peking Union Medical College, Beijing 100021, China*

Correspondence should be addressed to Xiabi Liu; liuxiabi@bit.edu.cn and Xiaohong Ma; dr_maxh_cams@sina.com

Mammography remains the most prevalent imaging tool for early breast cancer screening. The language used to describe abnormalities in mammographic reports is based on the Breast Imaging Reporting and Data System (BI-RADS). Assigning a correct BI-RADS category to each examined mammogram is a strenuous and challenging task for even experts. This paper proposes a new and effective computer-aided diagnosis (CAD) system to classify mammographic masses into four assessment categories in BI-RADS. The mass regions are first enhanced by means of histogram equalization and then semiautomatically segmented based on the region growing technique. A total of 130 handcrafted BI-RADS features are then extracted from the shape, margin, and density of each mass, together with the mass size and the patient's age, as mentioned in BI-RADS mammography. Then, a modified feature selection method based on the genetic algorithm (GA) is proposed to select the most clinically significant BI-RADS features. Finally, a back-propagation neural network (BPN) is employed for classification, and its accuracy is used as the fitness in GA. A set of 500 mammogram images from the digital database for screening mammography (DDSM) is used for evaluation. Our system achieves classification accuracy, positive predictive value, negative predictive value, and Matthews correlation coefficient of 84.5%, 84.4%, 94.8%, and 79.3%, respectively. To our best knowledge, this is the best current result for BI-RADS classification of breast masses in mammography, which makes the proposed system promising to support radiologists for deciding proper patient management based on the automatically assigned BI-RADS categories.

## 1. Introduction

Breast cancer is the most invasive and deadliest cancer in women worldwide. Recent statistical reports from the International Agency for Research on Cancer reported as many as 8.2 million deaths from cancer worldwide in 2012, and breast cancer ranked second after lung cancer with an incidence rate of 1.67 million [1]. Moreover, around 252.710 US women got breast cancer, and 40.610 of deaths have been expected in 2017 [2]. In China, breast cancer becomes the type of cancer most commonly diagnosed among women; it accounts for 12.2% of global cases and 9.6% of all deaths worldwide [3]. However, if diagnosed as early as possible, breast cancer can be substantially curable, which can greatly help to provide more treatment options and thus improve survival rates. Currently, X-ray screening mammography is a commonly used tool that plays a key role in the identification of early breast cancer and helps to reduce its mortality. Based on several factors such as radiologist training and dexterity, breast tissue density, and mammogram quality, it was observed that the mammography efficiency varies from 60 to 90% [4]. Likewise, it was reported that radiologists may overlook around 10%-30% of all mammogram lesions [5].

Furthermore, during the screening process, radiologists often visualize mammograms to look for the most common symptoms indicating the presence of cancer in breast tissue, i.e., mass, calcification, asymmetry, or architectural distortion [6]. They routinely analyze mammograms by referring to BI-RADS (Breast Imaging Reporting and Data System) [7] which is a standardization and quality assurance lexicon for mammographic reports developed by the American College of Radiology (ACR). The aim was to homogenize mammographic language between radiologists and referring clinicians and make it more clear and consistent. Besides, BI-RADS mammography encompasses qualitative features to characterize the mass shape, margin, and density. Then, depending on these features, the radiologists assign the mass lesion to one BI-RADS category from the following:

(i) Category 0: incomplete, further imaging evaluations are required

(ii) Category 1: negative, no abnormality found

(iii) Category 2: benign

(iv) Category 3: probably benign

(v) Category 4: suspicious finding

(vi) Category 5: highly suggestive of malignancy

(vii) Category 6: known biopsy-proven malignancy

Accordingly, radiologists recommend an annual screening for categories 1 and 2, a six-month follow-up for category 3, and a biopsy for categories 4 and 5 [7]. However, radiologists have to read and interpret a tremendous number of mammograms daily, which is a repetitive, arduous, and error-prone process. Meanwhile, meticulously assigning a BI-RADS category to each examined mammogram is a laborious and challenging task even for experts. Consequently, a wide inter-observer variability when applying BI-RADS lexicon has been documented which often lead to classification errors. Boyer and Canale [8] have studied the reasons and categorized the interpretation errors made by mammographers into excess errors and default errors; the former occurs when a benign lesion is wrongly graded as suspicious (BI-RADS category 4 or 5), and the latter is encountered when radiologists misclassify a suspicious abnormality as benign or probably benign (BI-RADS category 2 or 3). Such reported errors would certainly have an adverse effect on the associated management recommendation reports and would have harmful consequences for the patient prognosis.

Recent advances in machine learning and image processing have culminated in the emergence of computer-aided diagnosis (CAD) systems which have been often used as an additional and useful tool to help doctors make final diagnostic decisions and act as a second opinion. Recently, a broad range of CAD systems have been proposed and achieved remarkable performance to predict breast cancer from mammography images [9–17]. However, most of these works focused on the classification of the detected breast abnormalities as either benign or malignant (i.e., pathology classes). From a clinical point of view, this binary classification is not congruent and disagrees with radiologist assessments. Indeed, in the clinical practice, radiologists should primarily put each lesion into one of the above-mentioned BI-RADS assessment categories, because they cannot claim the malignancy or benignity of the detected abnormality without carrying out a follow-up study or biopsy [18]. In addition, the entire diagnostic process for classification of mass lesions using CAD systems usually involves several stages: preprocessing, segmentation, feature extraction, feature selection, and classification. Feature selection is a very important stage to enhance the classification performance. Genetic algorithm- (GA-) based feature selection has been widely used in CAD-based breast cancer diagnosis and has shown remarkable efficacy to improve overall performance. For example, in [10, 11], a 10 and 20 GA-based selected features have been found to be efficacious for the classification of benign and malignant masses. However, as far as we are aware, there has been no previous study that endeavored to analyze the effect of feature selection on the BI-RADS-based mammogram mass classification performance.

This paper proposes a new and effective CAD system for the classification of mammographic masses into four categories in BI-RADS, including benign (B-2), probably benign (B-3), suspicious finding (B-4), and highly suggestive of malignancy (B-5). As we focus solely on the classification of detected masses, the categories (0, 1, and 6) are not considered in the proposed system. As an example, Figure 1 shows four mammogram images taken from the digital database for screening mammography (DDSM), which are used in our work. The suspicious regions bounded by the red rectangles are reflected in the upper middle of each mammogram image. From left to right, the mammograms are clinically assessed as B-2, B3, B-4, and B-5, respectively. For the discrimination of these four categories, our proposed system involves suitable pipeline stages: preprocessing, segmentation, feature extraction, feature selection, and classification. First, the mass regions are preprocessed and enhanced using histogram equalization. Second, a semiautomatic segmentation method is introduced based on the region growing technique to isolate the masses from the neighboring breast tissues. Third, we extract a set of handcrafted features from the mass shape, margin, and density, together with the mass size and the patient's age, as stated in BI-RADS mammography. We then propose a modified GA-based feature selection method for selecting the most clinically representative BI-RADS features and demonstrating its significant impact on classification. This method is designed in a customized fashion with problem-specific operators to identify the best feature subsets corresponding to each number of features that we explore successively. Among all of these subsets, the best feature subset is then obtained as the optimal for the considered classification. Finally, the classification is carried out using a BPN classifier whose accuracy is employed as the fitness in GA.

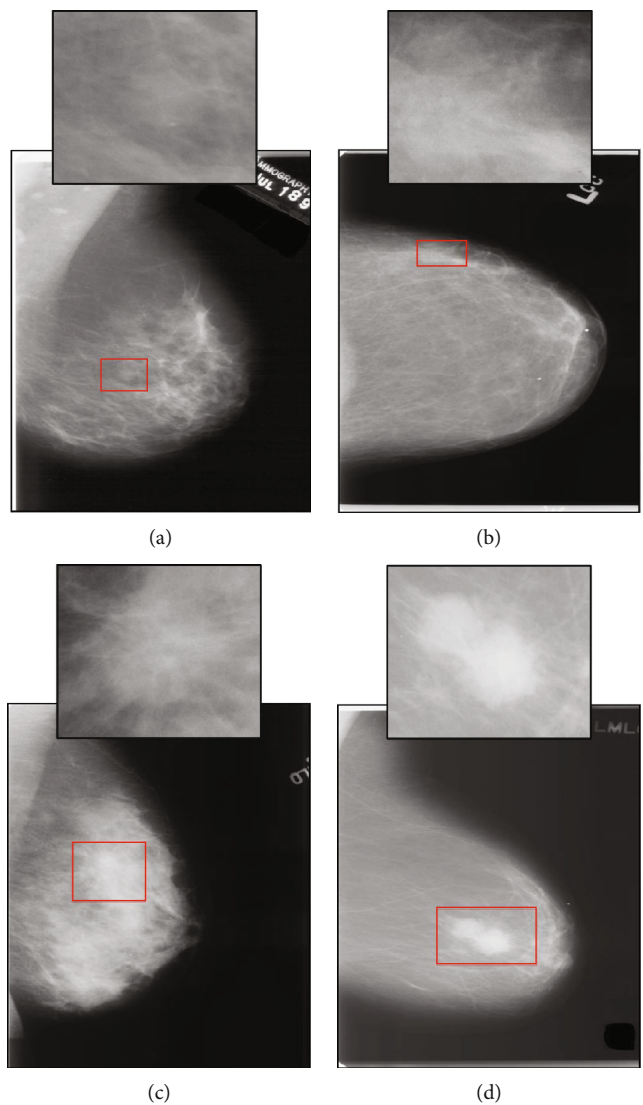

Figure 1: Mammogram samples of the four BI-RADS categories taken from the DDSM database: (a) B-2 (A_2001_1.RIGHT_MLO), (b) B-3 (B_3099_1.LEFT_CC), (c) B-4 (B_3390_1.LEFT_CC), and (d) B-5 (C_0176_1.LEFT_MLO). The extracted regions of interest (ROIs) are shown in the upper middle of each image.

The main contributions of this paper are summarized as follows:

(i) We develop a new and effective CAD system aimed at classifying an input mammographic mass into one of four assessment categories in BI-RADS (B-2, B-3, B-4, and B-5). To our best knowledge, it outperforms the state-of-the-art counterparts and achieves the best current performance of BI-RADS breast mass classification in mammography

(ii) In order to ensure more accurate segmentation results, we introduce a semiautomatic segmentation method based on the region growing technique to separate mass lesions from surrounding breast tissues

(iii) To improve the classification performance, we propose a modified GA-based feature selection method where we first look for the best feature subset from each number of features we explore; then, the optimal subset among them is deduced for the classification. For the first time, we analyze and demonstrate the efficiency of feature selection on the BI-RADS classification based on the proposed method

The rest of this paper is structured as follows. In Section 2, we review the related works. Section 3 details the proposed CAD system. In Section 4, the obtained results are presented and discussed. Section 5 concludes this study and drives some future directions and perspectives.

## 2. Related Works

In this section, we analyze briefly some existing CAD systems for classifying breast cancer from mammography images, which we found related to our proposed approach.

Since the first release of BI-RADS, the use of the BI-RADS mammography atlas in the research community has been restricted either to exploit the included features of findings for detection and/or classification tasks or to study and evaluate the inter-observer variability between radiologists when applying the lexicon [8, 19, 20]. In [5], the authors introduced a CAD system based on combined handcrafted BI-RADS features from two mammographic views (Mediolateral-Oblique (MLO) and Cranio-caudal (CC)) of 115 images from the DDSM database. They passed the mass shape and margin as well as the patient's age features as inputs to the Linear Discriminant Analysis classifier and achieved an area under the curve (AUC) of 0.92. Surendiran and Vadivel [21] utilized 17 geometrical and margin features to classify mammographic masses as benign, malignant, or normal. They used 1553 DDSM masses with the Classification and Regression Tree classifier and achieved an overall accuracy of 93.72% for benign and malignant classifications and 95.68% for ternary classification (benign vs. malignant vs. normal). In their subsequent study [22], the authors evaluated the effects of BI-RADS features on the classification of breast mammograms; they introduced 20 multimodal features consisting of 17 handcrafted BI-RADS features representing the shape, texture, and margin as well as 3 DDSM database descriptors including assessment, subtlety, and density for benign and malignant mass classification. They trained a univariate ANOVA discriminant analysis classifier using a total of 300 DDSM mammograms. The results obtained have demonstrated that using the 20 combined sets of features including DDSM descriptors yielded better accuracy (93.3%) compared to individually used quantitative BI-RADS features (86.7%). In [23], a method for classifying normal and abnormal patterns from mammograms has been introduced. Images obtained from the DDSM database were used. The authors described the mammogram texture using 5 extracted GLCM features. By using a neural network classifier, they obtained a maximum accuracy rate of 96%. Rabidas et al. [13] proposed two new feature extraction methods based on neighborhood structural similarity for the characterization of mammographic masses as benign or malignant. The neighborhood similarity-based features are further combined with local binary pattern-based features to enhance the performance. Stepwise logistic regression-based feature selection was performed to extract the optimal feature subset which was then fed as input into Fisher linear discriminant analysis for classification. The authors reported the accuracies of 94.57% and 85.42% using the mini-MIAS and DDSM databases, respectively. Vadivel and Surendiran [24] extracted 17 margin and shape features to classify mammographic masses into BI-RADS shape categories: oval, round, lobular, and irregular. Their experiments were performed using 224 DDSM masses, and a C5.0 decision tree classifier correlated with a fuzzy inference system was used to drive the classification. Their results were encouraging compared with a similar proposed method. These research studies have demonstrated the effectiveness of using handcrafted BI-RADS features derived from mammographic lesions for the diagnosis of breast cancer. Generally, the more features extracted, the higher the prediction accuracy will be. However, with this hypothesis, the model used becomes difficult to be interpreted and more likely to overfit, especially if the available data is relatively small. For this reason, researchers shifted the direction towards introducing a feature selection stage before the classifier in order to reduce the number of collected features while preserving the classification accuracy. GA-based feature selection has been widely used and showed efficiency to improve the performance. Elfarra and Abuhaiba [10] used the GA to improve the classification accuracy of 410 DDSM mammograms into benign, malignant, and normal. Their GA selected 10 discriminative features out of 65, which are then fed along with other features to the SVM classifier. Rouhi et al. [11, 25] used GA to reduce the feature space to a 20-element feature vector before feeding it to different classifiers in order to evaluate their proposed segmentation techniques. Deep learning-based approaches, especially with a convolutional neural network (CNN), have been also introduced in recent years for mammogram breast cancer classification and reported remarkable results [15, 16, 26].

However, the above-mentioned works focused on discriminating mammogram benign lesions from malignant or normal from abnormal, while few studies have reported the BI-RADS classification of breast masses using mammograms. The authors in [18] introduced an automatic CAD system to classify mammographic masses either as benign or as malignant or in four BI-RADS categories (B-2, B-3, B-4, and B-5). A set of 23 handcrafted features was extracted and fed into a multilayer perceptron for classification. They used 480 DDSM masses to evaluate their approach. They achieved 88.02% and 83.85% accuracies for binary classification (benign vs. malignant) and multiclass classification (BI-RADS classification), respectively. However, the effect of the handcrafted features on classification has not been highlighted, and, to our point of view, the final feature subset used in the authors' experiments has most likely been selected on the basis of some prior knowledge from clinical experience or on a trial-and-error strategy. The authors in [27] proposed a CAD system based on fuzzy logic concepts to represent the image features and categorize the mammographic lesions (mass and calcification) into BI-RADS classes. They asked 5 radiologists to evaluate the mass shape, margin, and density and used the resulting attributes as inputs to a fuzzy inference system to predict four BI-RADS categories (B2, B-3, B-4, and B-5). Their fuzzy BI-RADS system was trained and tested using 46 mammogram masses from the DDSM database, and an accuracy of 76.67% was obtained. However, due to the few number of mammograms used, their method is not highly guaranteed. In addition, human experts have performed the evaluation of abnormalities, making the process vulnerable to subjectivity, easily affected by customized analysis, and often leading to discrepancies. Recently, Domingues et al. [28] proposed two new preprocessing techniques based on data augmentation and multiscale enhancement in order to classify mammograms into BI-RADS classes. For feature extraction, the authors investigated a pretrained CNN with the AlexNet model [29]. Their experiments were performed on the publicly available INbreast dataset [30]. The best classification result achieved based on the proposed techniques was 83.4%.

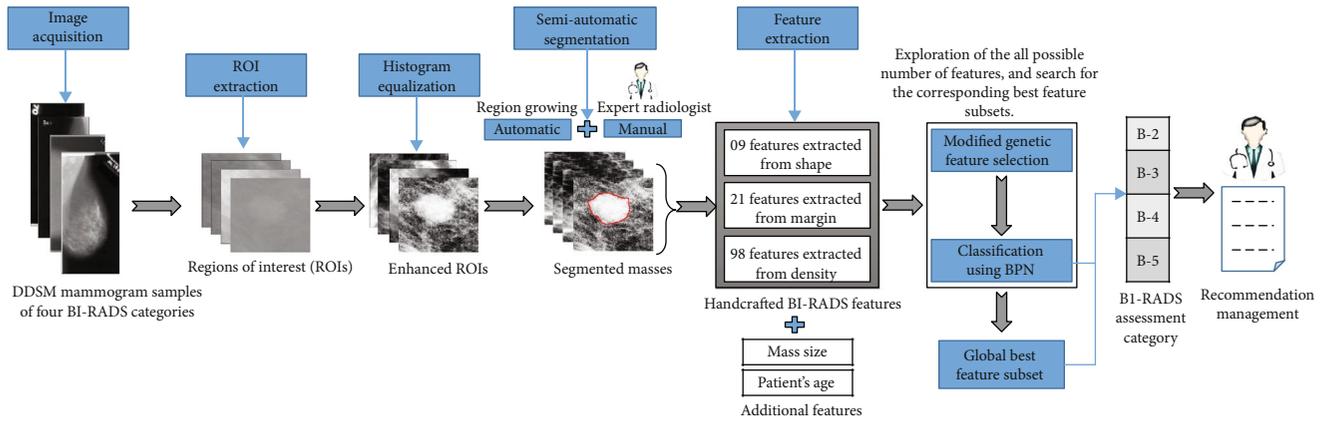

Figure 2: Block diagram of the proposed CAD system.

However, the CNN-based approaches require a massive amount of data to train efficiently. In addition, they learn global representations from the global image area, which are usually difficult to interpret. Moreover, some important features for the BI-RADS classification, such as the patient's age, cannot be directly extracted using a CNN. Our approach, on the contrary, has the advantage of identifying the most clinically significant BI-RADS features for diagnosis.

## 3. Methodology

As described above, this paper proposes a new CAD system aimed at differentiating between four BI-RADS assessment categories (B-2, B-3, B-4, and B-5) from digitized mammograms. Figure 2 illustrates the block diagram of the proposed system that involves five main stages: preprocessing, semiautomatic segmentation of regions of interest (ROIs), feature extraction, feature selection, and classification. They are detailed below.

*3.1. Mammogram Image Preprocessing.* First, the information provided in the dataset (the center and the approximate radius of each abnormal area) is used to crop the full mammogram images and get the corresponding ROIs. Since mammograms often appear with artifacts, noise, and weak contrasts, a technique for image enhancement should be applied to reduce the noise and increase the contrast between the lesion and its surrounding tissue, which is important for further image analysis. In our work, we used the histogram equalization (HE) to spread out the most frequent intensity values across the total range in order to achieve higher contrast. HE expands the dynamic range of the input image histogram based on its cumulative probability distribution function. Consequently, a uniform distribution of grey levels is obtained, and the mammogram information is optimized. Figure 3 illustrates an example of the enhancement obtained by applying HE on a mammogram mass assessed as probably benign (B-3). It is clear that the HE provides better visualization of the mammogram ROI, and more useful image features can be easily observed.

*3.2. Segmentation.* In this stage, the obtained ROIs in the preprocessing step are segmented to separate the mass area from its neighboring breast tissues. Due to the complex morphology of mass lesions that often appear with ambiguous and jagged contours and the nature of their surrounding fibroglandular breast tissue, the segmentation of masses is much more difficult compared with other findings such as calcification or architectural distortion. Several segmentation methods have been proposed in the literature to extract the precise contour of mammographic lesions which are roughly categorized into automatic [11, 31, 32], semiautomatic [33, 34], and manually performed by the radiologist [18, 35]. However, previous studies [36] have shown that there is no "one-fit-all" segmentation technique and none of the automatic and manual methods can provide fully accurate results for diverse mass lesions in mammograms. The reason is that manual (hand-drawn) segmentation can eliminate useful details of the contour, especially for malignant masses, and on the other hand, automatic segmentation cannot always yield satisfactory accuracy due to the complex nature of masses involving various textures, sizes, and fuzzy boundaries. To deal with these issues and to ensure good segmentation results, we introduce a combination of manual and automatic approaches to semiautomatically segment the ROIs.

Region growing (RG) is an automatic and widely used region-based segmentation method, which consists of selecting random pixels in the image and then merging them to neighbors if they are homogeneous; otherwise, they are labeled as boundary pixels. We first use the RG technique to get the initial mass contours. The seed point from where the region starts growth is defined as the center pixel of each ROI image. However, an optimal threshold to stop the region's growth for all images is often hard to find. The appropriate threshold value is usually searched experimentally to avoid under and oversegmentation. Here, we defined a range of values bounded by the minimum and the maximum grey levels of each ROI image. By this way, different grown regions of a single mass image are obtained with different thresholds. Based on each ROI image, an expert radiologist in mammography is asked first to choose the most appropriate thresholds up to his experience, then select and refine the most precise mass contour from the obtained grown regions to get the final segmentation result. Figure 4 shows the final obtained segmented ROIs of four mammogram images assessed as B-2, B-3, B-4, and B-5 by using the proposed semiautomatic segmentation method.

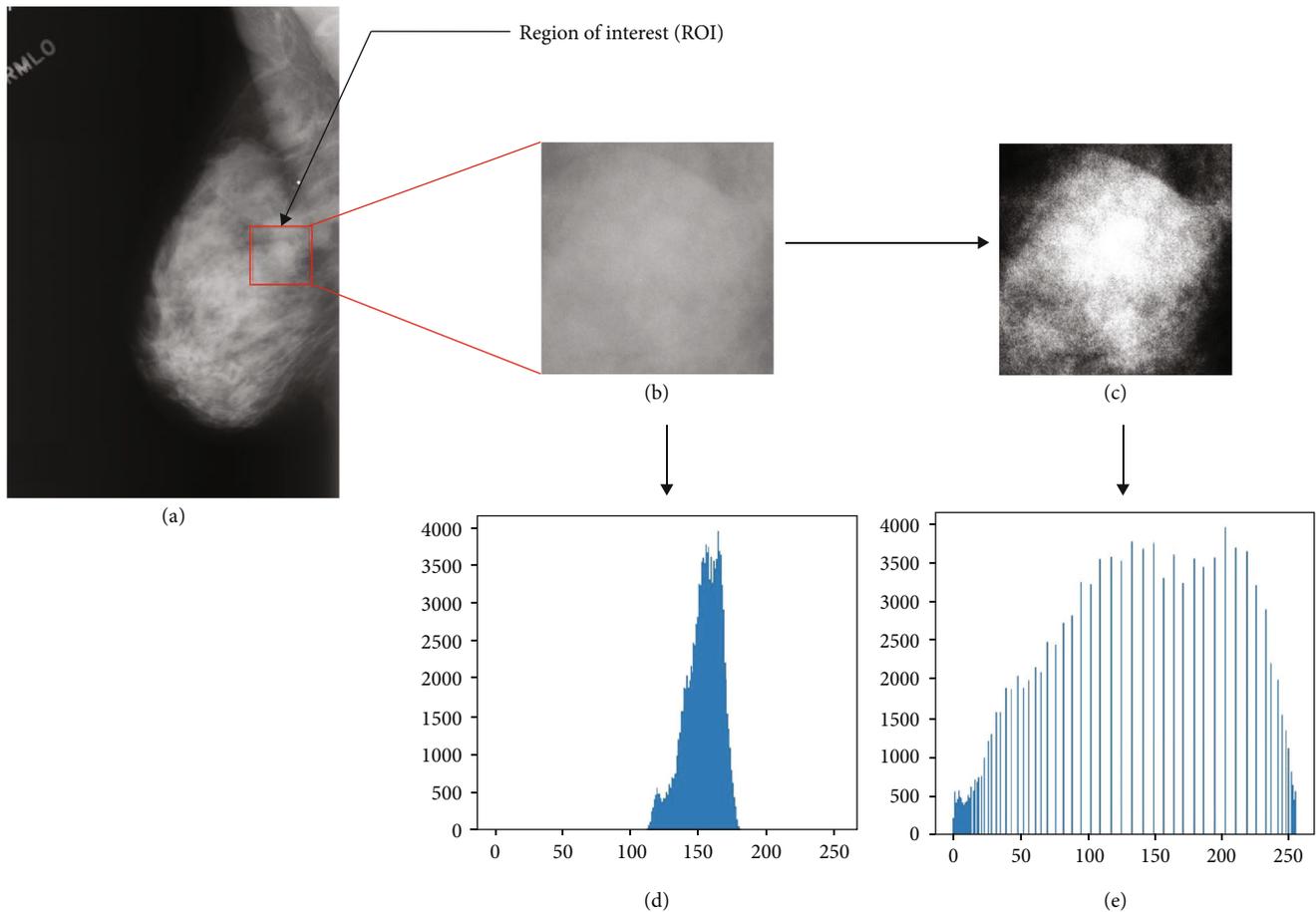

Figure 3: Preprocessing stage: (a) full mammogram image, (b) ROI obtained by cropping (a), (c) final obtained ROI after applying HE on (b), (d) original histogram, and (e) enhanced histogram.

3.3. Feature Extraction. In this section, a set of handcrafted BI-RADS features is quantified from the segmented ROIs. According to the BI-RADS atlas [7], a mass is "a space-occupying lesion seen in two different projections" and it can be distinguished by its morphological and textural features; the former includes shape and margin characteristics, while the latter describes the mass density. Besides, as indicated in Figure 5, the BI-RADS depicts the mass shape as *round*, *oval*, *lobular*, or *irregular* and the margin as *circumscribed*, *microlobulated*, *obscured*, *indistinct* (*ill-defined*), or *speculated*. Lastly, the density is characterized with respect to the breast glandular tissue as *low density*, *equal density*, *high density*, or *fat-containing radiolucent*. Generally, these descriptors are arranged from the lowest indicative of malignancy to the highest. Masses with round or oval shape, circumscribed margin, and/or low density and fat-containing are more prospective to be benign lesions (mostly B-2 or B-3 in clinical terms), while masses with irregular shape, speculated margin, and high density have high likelihood of being tumorous (mostly B-4 or B-5 in clinical terms) [37].

To cover the entire categories mentioned in BI-RADS mammography, we extracted a total of 130 features (9 features from shape, 21 from margin, 98 from density, and the 2 additional features that represent the mass size and the patient's age). These features and their computing methods are represented below in detail.

3.3.1. Shape Features. In the clinical practice, radiologists often use the shape descriptors to estimate the suspicion level of the detected masses in mammograms. A breast mass can be of round shape, oval, lobular, or irregular. In our work, we used different features to represent the shape. First, by using the active contour method [39], we extracted three features, namely, *continuity*, *curvature*, and *irregularity*. They were also used in [18]. Second, we investigated six other features introduced in [40] to distinguish benign masses from malignant on sonography, including *difference area* (*convex hull area minus mass area*), *mean variation*, *variance variation*, *skewness variation*, *kurtosis variation*, and *entropy variation*. As a result, in total, we quantified 9 BI-RADS features to describe the mass shape.

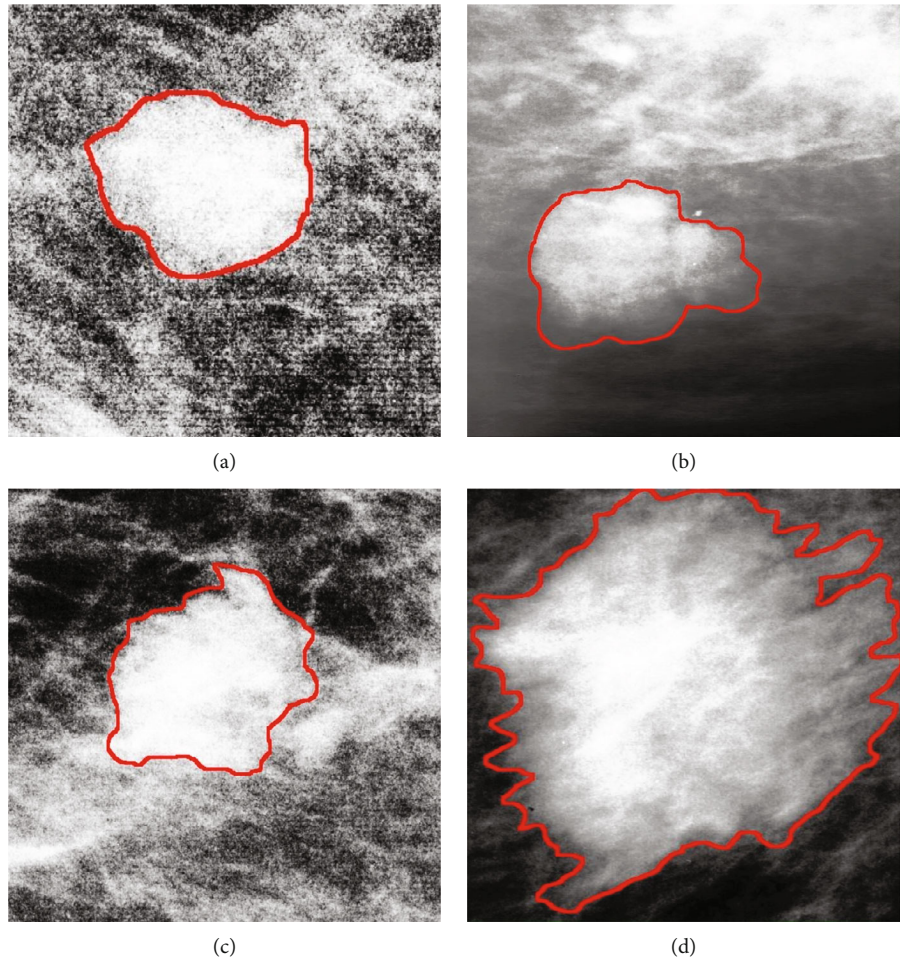

(a)  (b)  (c)  (d)

Figure 4: Final segmented ROIs after applying HE and the proposed semiautomatic segmentation method: (a) B-2 sample, (b) B-3 sample, (c) B-4 sample, and (d) B-5 sample.

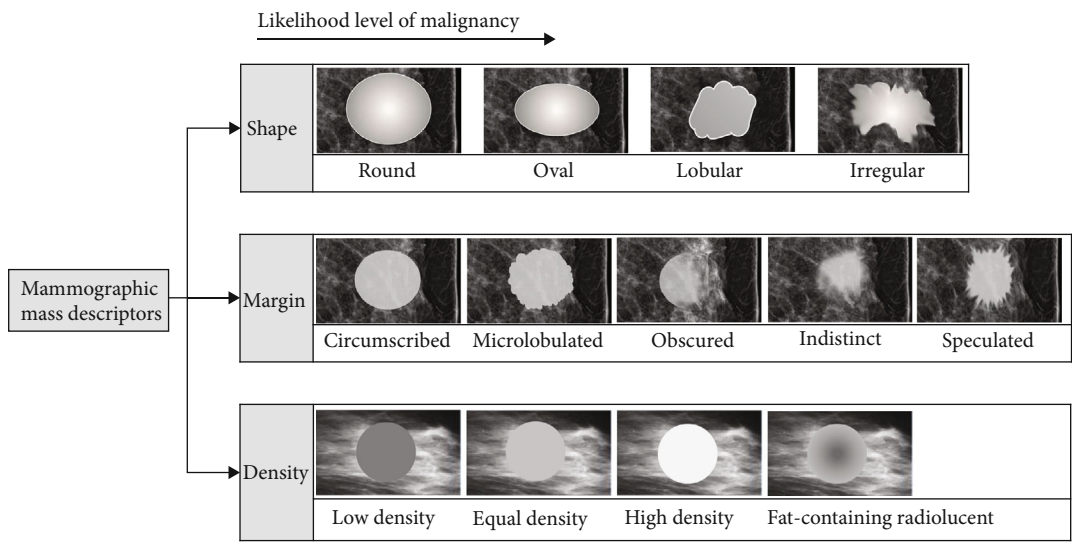

Figure 5: Mass lesion descriptors according to BI-RADS mammography [38].

*3.3.2. Margin Features.* As it was reported in BI-RADS mammography [7], the margin characteristics are essential for the differentiation between the different assessment categories. For the extraction of margin features, a set of waveforms are used in our study, which are sequentially selected while traversing along over the margin of the mass. After that, an edge probability vector is calculated for each waveform. The main advantage of using a waveform is to catch the abruptness within and outside the mass. Hence, we followed the work of Bagheri et al. [41] to extract *kurtosis*, *entropy*, and *index of the maximum probability* as three main margin features. However, it is important for a fixed number of waveforms along the margin to have normalized margin features for different mass sizes identified in mammograms. Thus, we set the number of waveforms to 32, the waveform length to 64, and the angle $\theta$ for the sequential interval to place a new waveform along the margin as $\pi/16$. More details of this process can be found in [18, 41]. To reduce the number of the extracted features, we used different statistical functions: mean, maximum, minimum, standard deviation, variance, skewness, and kurtosis for each of the three extracted features over 32 waveforms. As a result, we obtained 21 handcrafted BI-RADS features describing the mass margin.

*3.3.3. Textural Features.* Here, we quantified 98 textural BI-RADS features by means of the Grey Level Cooccurrence Matrix (GLCM) [42]. GLCM relies on extracting second-order statistical features by modeling the relationship between groups of two pixels within the same region and measuring the appearance frequency of their corresponding grey levels that are separated by an angle $\alpha$ and a distance $d$. Because of wide varieties in mammogram tissues, tumors often appear with different texture scales (coarse, fine, etc.). For this reason and to obtain a better representation of interpixel correlations, we followed the work of [10] at choosing multiple distances $d$ and eight angles $\alpha$, where the cooccurrence matrix for a square ROI is calculated using $\alpha = \{0, \pi/8, \pi/4, 3\pi/8, \pi/2, 5\pi/8, 3\pi/4, \text{and } 7\pi/8\}$, and at all distances $d \in \{0, 1, 2, 3, ..., L/2\}$, where $L$ is the length of a ROI's side. Unlike the work of [10], we quantified all the 14 GLCM features: *angular second moment*, *contrast*, *correlation*, *variance (sum of squares)*, *inverse difference moment*, *sum average*, *sum variance*, *sum entropy*, *entropy*, *difference variance*, *difference entropy*, *information measure of correlation 1*, *information measure of correlation 2*, and *maximal correlation coefficient*. However, the same as the margin features, the quantification of 14 GLCM features for each value of $d \in [1, L/2]$ will lead to a high dimensional feature vector; therefore, we employed the same seven statistical functions mentioned in margin features for each cooccurrence feature over all the values of $d$. As a result, we obtained a feature vector of $14 \times 7 = 98$ elements describing the texture of each segmented mass.

*3.3.4. Additional Features.* In addition to the mass size, the BI-RADS mammography [7] reported that the patient's age is necessary for the categorization of mammograms as it has the ability to change the assessment category of the lesion (downgrade or upgrade the BI-RADS category). This issue was also formerly discussed in the works of Lo et al. [43], Gupta et al. [5], and Chokri and Farida [18], in which the age has been found as an important discriminatory feature for the BI-RADS-based CAD systems.

Table 1 provides a summary of the aforementioned quantified BI-RADS features. ∗ marked in the margin and density categories indicates that seven statistical functions have been assigned to each feature in those categories: mean, maximum, minimum, standard deviation, variance, skewness, and kurtosis.

After quantifying the above-mentioned features, we found a significant range variance between data. Thus, to prevent certain features from dominating others, we normalized the values of features within the range [0, 1]. To do so, we applied the min–max feature scaling technique following this formula:

$$d_{\text{norm}} = \frac{d - d_{\min}}{d_{\max} - d_{\min}}, \quad (1)$$

where $d$ is the original feature value and $d_{\text{norm}}$ is its normalized value.

*3.4. Modified Genetic Feature Selection.* Over the last few decades, many feature extraction methods have been introduced to glean handcrafted features from mammography images which have been further used to build effective CAD systems for breast cancer diagnosis. In fact, not all the features collected are equally important and do not have the same capacity for discrimination, some of which are redundant and may carry irrelevant information. Such features may contribute to the deterioration of the classifier performance. Contrarily, a subset of appropriate features can be sufficient to provide higher and robust classification performance [44]. In this work, we apply the genetic algorithm (GA) to search the best subsets of features and eliminate the insignificant features.

In the previous methods of GA-based feature selection, the chromosomes are encoded with genes; each gene is a bit, which takes a value of either 1 or 0. Each gene position in the chromosome corresponds to a specific feature. Thus, the optimization process evolves chromosomes containing all the available features, and a feature is marked "selected" for the solution of the classification problem if its corresponding bit value at the gene position is 1. Otherwise, it is marked "discarded." This strategy could drop important candidate solutions and also slow down the optimization process. Differently, we assign each BI-RADS feature to a gene, so the length of the chromosomes ($L$) is the number of considered features. Based on such representation, we design GA in a tailored fashion with problem-specific operators to test all the potential solutions and figure out the best feature subsets corresponding to each number of features that we explore successively. Then, the global optimal subset among all those subsets is obtained for the considered classification.

Since we have 130 BI-RADS features, we start the exploration successively from $L = 1$ until we reach $L = 130$. This led us to test every one BI-RADS feature separately when $L = 1$ and test the entire feature space when $L = 130$, which means that no feature selection is involved; also, we performed GA 128 times in succession and independently for each number of features explored when $L$ ranges from 2 to 129. This schema is illustrated in Figure 6. As shown, GA only needs to be applied when $L$ ranges from 2 to 129, the steps in which are the following:

Table 1: Summary of all quantified BI-RADS features used in this study.

| BI-RADS category | Descriptors | Feature number | Feature name |
| --- | --- | --- | --- |
| Shape | Round<br>Oval<br>Lobular<br>Irregular | 1-9 (9 features) | Continuity<br>Curvature<br>Irregularity<br>Difference area: convex hull area, minus mass area<br>Mean variation<br>Variance variation<br>Skewness variation<br>Kurtosis variation<br>Entropy variation |
| Margin* | Circumscribed<br>Microlobulated<br>Obscured<br>Indistinct (ill-defined)<br>Speculated | 12-32 (21 features) | Kurtosis<br>Entropy<br>Index of the maximum probability |
| Density* | High<br>Equal<br>Low<br>Fat-containing radiolucent | 33-130 (98 features) | Angular second moment<br>Contrast<br>Correlation<br>Variance (sum of squares)<br>Inverse difference moment<br>Sum average<br>Sum variance<br>Sum entropy<br>Entropy<br>Difference variance<br>Difference entropy<br>Information measure of correlation 1<br>Information measure of correlation 2<br>Maximal correlation coefficient |
| Additional features | — | 10-11 (2 features) | Mass size<br>Patient age |

*Step 1.* Initialization: creating an initial population with $N$ number of chromosomes.

*Step 2.* Chromosome evaluation: evaluating each individual or chromosome in the population using the objective or fitness function.

*Step 3.* Reproduction: creating a new population of $N$ chromosomes by using genetic operators (selection, crossover, and mutation) on certain chromosomes in the current population.

*Step 4.* Loop: back to Step 2 as long as the problem stopping criterion is not satisfied.

In the following, we provide the details of our modified GA-based feature selection algorithm.

*3.4.1. Encoding Scheme.* Encoding has the key role to play in turning the solution of the problem into a chromosome. In our work, each gene in the chromosome corresponds to a BI-RADS feature; then, we assign a value ranging from 1 to 130 for each gene as an *identifier* (*id*) for one out of 130 extracted features. By this way, we arranged the features on a specific ascendant order from 1 to 130, the same as illustrated in Table 1.

*3.4.2. Initialization.* Here, an initial population with $N$ number of chromosomes is created. We set the size of the populations to a variable number multiplied by 4, and we seeded them with a random initialization. The number of generations and the number of chromosomes in each generation were determined on the basis of the size of the corresponding feature vector used, i.e., the length of the chromosome. Since the length of chromosomes in our method is increased sequentially after each explored number of features, we slowly reduce the number of generations and the number of chromosomes in each generation, in order to achieve a good balance between computational time and fast convergence.

*3.4.3. Fitness Function.* This function helps determine how well a candidate chromosome is able to solve the problem. Since we applied GA separately, we used only the classification accuracy of the back-propagation neural network (BPN) as a fitness function to select the best chromosomes in each generation in order to breed the next generation.

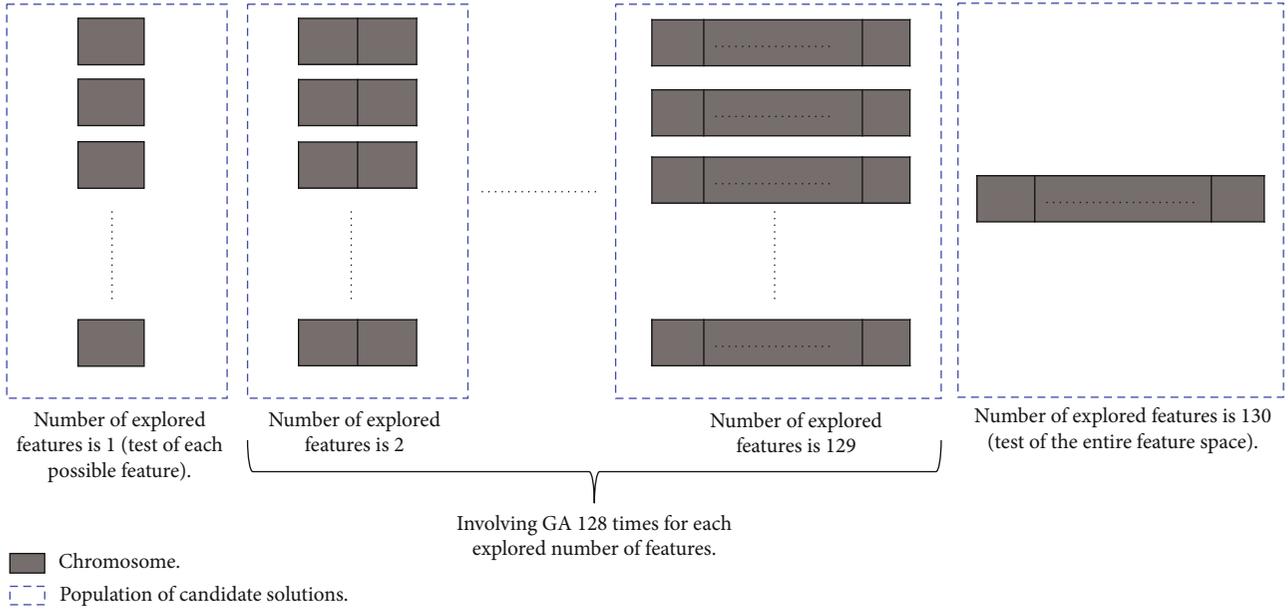

Figure 6: General architecture of the modified GA-based feature selection method.

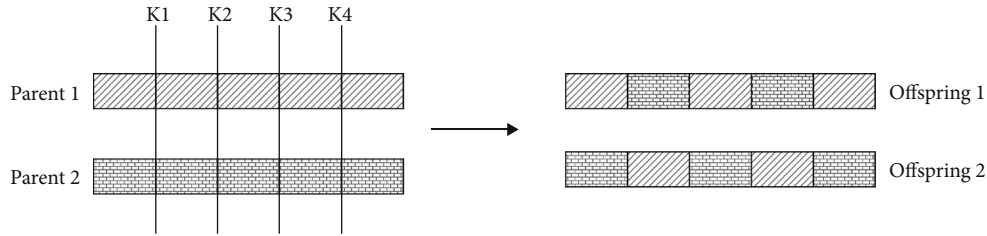

Figure 7: Example of the performed crossover operator.

*3.4.4. Selection Operator.* With the selection operator, individuals or parents of the current population are selected to survive, mate, and create a new population. In our algorithm, the selection of the parents to mate and recombine for creating offspring for the next generations is based on two strategies: firstly, we exploited the *elitism* mechanism to select the best two chromosomes in the current population to be always propagated and directly copied into the next generation. Secondly, the remaining chromosomes were chosen using the well-known *roulette wheel* selection method [45]. This is dependent on their fitness score, in which the fittest chromosomes are more likely to reproduce. Thereby, a chromosome is selected probabilistically; its selection probability $P(c_i)$ is calculated by using this formula:

$$P(c_i) = \frac{f(c_i)}{\sum_{j=1}^{k} f(c_j)}, \quad (2)$$

where $c_i$ is the current chromosome, $f(c_i)$ is its corresponding fitness score, and $k$ is the population size. This combination of *elitism* and *roulette wheel* selection methods can significantly ensure the optimum global convergence of GA. This is justified by the fact that, by using the *elitism* mechanism, we are extremely wary of losing the best candidate solutions that can often occur while generating a new generation using mutation and crossover operators. Moreover, with the *roulette wheel* method, each of the remaining chromosomes will have chance to be selected in order to breed the next generations.

*3.4.5. Crossover Operator.* This operator is used to create new child chromosomes by recombining the genes of two selected parents. Here, we adopted the $K$-point crossover technique to our algorithm with a defined probability $P(c) = 1$. The $K$-point crossover selects more than one crossover point to produce the offspring. Here, we subtracted 1 from the length $L$ of the parent chromosomes to get the number of crossover points $K$. Then, the genes are exchanged one by one progressively until reaching the maximum crossover points. An example of the performed $K$-point crossover is shown in Figure 7 (in this example, $K = 4$ and $L = 5$).

*3.4.6. Mutation Operator.* After the crossover is accomplished, the mutation operator will take place. Based on the fact that each gene of our chromosomes corresponds to one different feature using an integer *id* ranging from 1 to 130, thus, if no repeated *id* is encountered after crossover, we used a random resetting technique for integer representation to introduce the mutation to our GA algorithm using a low probability $P_m = 0.01$; otherwise, we introduce a conditional and guided mutation to replace only the repeated *id* with a new one that is randomly generated from the set {1...,130}. This mutation strategy has the main benefits of omitting redundant features to be fed as inputs to the BPN classifier and preserving the variety of solutions within the population.

*3.4.7. Termination Condition.* In our study, the GA will stop when no improvement in the fitness scores is noticeable for $x$ iterations.

Note again that the main difference between our GA-based feature selection method above and the previous counterparts lies in the whole process of the GA-based method and, in particular, in how chromosomes are encoded with the extracted features, how the feature subsets are explored, and how the convergence of optimizations is achieved through problem-specific operators.

*3.5. Classification.* The BPN is employed as the output classifier to predict the four BI-RADS categories based on the input features, which consists of an input layer, one hidden layer, and an output layer. The number of neurons in the output layer is fixed to 4, representing the four BI-RADS categories, while the number of neurons in the input layer varies with the number of features currently being explored. As for the number of neurons in the hidden layer, let it be denoted as $H$; we follow a rule described in [18] to determine it:

$$H = O + (0.75 \times I),$$
$$H < 2 \times I, \quad (3)$$

where $I$ is the number of input features and $O = 4$. For example, in our experiments, we have $H = 1$ for $I = 1$, $H = 101$ for $I = 130$, $H = 38$ for $I = 46$, etc.

## 4. Experiments

*4.1. Data Acquisition.* A set of 500 mammograms taken from the digital database for screening mammography (DDSM) is used to build and evaluate our proposed CAD system [46]. The DDSM is a public database created by the University of South Florida for use in the research community and is freely available at (*DDSM*:http://www.eng.usf.edu/cvprg/Mammography/Database.html). It contains approximately 2620 studies divided into normal, benign, and cancer cases. Every case contains two images of each breast from two different views: MLO and CC. Mammograms in DDSM were scanned with a resolution between 42 and 100 microns; the location of each abnormality is also provided. Moreover, some related patient and image information was included such as the date of the study, patient's age, subtlety class for lesions, BI-RADS density assessment, the final pathology result, and the BI-RADS assessment categories. It is worth noting that we focus only on pure mass lesions and exclude the masses with calcifications or architectural distortions, and we use images from both views as they can provide better classification results such as in [5]. To train and test the proposed model, 60% of the available set is allocated for training and the remaining 40% is used for testing.

Table 2: Confusion matrix of the best feature subset for classification.

| BI-RADS categories | Predicted | | | |
|---|---|---|---|---|
| Actual | B-2 | B-3 | B-4 | B-5 |
| B-2 | **47** | 2 | 1 | 0 |
| B-3 | 3 | **41** | 4 | 2 |
| B-4 | 2 | 5 | **36** | 7 |
| B-5 | 0 | 3 | 2 | **45** |

*4.2. Evaluation Criterions.* We use the confusion matrix, the accuracy, the sensitivity and the specificity, the positive predictive value (PPV), the negative predictive value (NPV), and Matthews correlation coefficient (MCC) to evaluate the classification performance. The confusion matrix contains information about the predicted and actual classes. The accuracy is defined as the proportion of the total number of correct predictions divided by the total number of testing data as follows:

$$\text{Accuracy} = \frac{TP_{B-2} + TP_{B-3} + TP_{B-4} + TP_{B-5}}{200}, \quad (4)$$

where $TP_{B-2}$, $TP_{B-3}$, $TP_{B-4}$, and $TP_{B-5}$ represent the true positive fractions of the four BI-RADS categories, respectively.

The sensitivity and specificity are used to measure the effectiveness of the classifier in correctly distinguishing each of the four BI-RADS categories. We also computed the PPV, NPV, and MCC as evaluation parameters. These metrics are determined as follows:

$$\text{Sensitivity} = \frac{TP}{TP + FN}, \quad (5)$$

$$\text{Specificity} = \frac{TN}{TN + FP}, \quad (6)$$

$$\text{PPV} = \frac{TP}{TP + FP}, \quad (7)$$

$$\text{NPV} = \frac{TN}{TN + FN}, \quad (8)$$

$$\text{MCC} = \frac{TP \times TN - FP \times FN}{\sqrt{(TP + FP)(TP + FN)(TN + FP)(TN + FN)}}, \quad (9)$$

where TP refers to the true positive fraction, TN refers to the true negative fraction, FN refers to the false negative fraction, and FP refers to the false positive fraction. For the above-mentioned criterions, higher values indicate better classification performance. The MCC is widely used in machine learning as an evaluation metric to assess the quality of classification. We instantiated the microaveraging technique to sum all the values of TP, TN, FP, and FN for each BI-RADS class and then compute PPV, NPV, and MCC.

*4.3. Results and Discussions.* Table 2 shows the confusion matrix of the best feature subset that produces the highest classification accuracy. The correct predictions are located

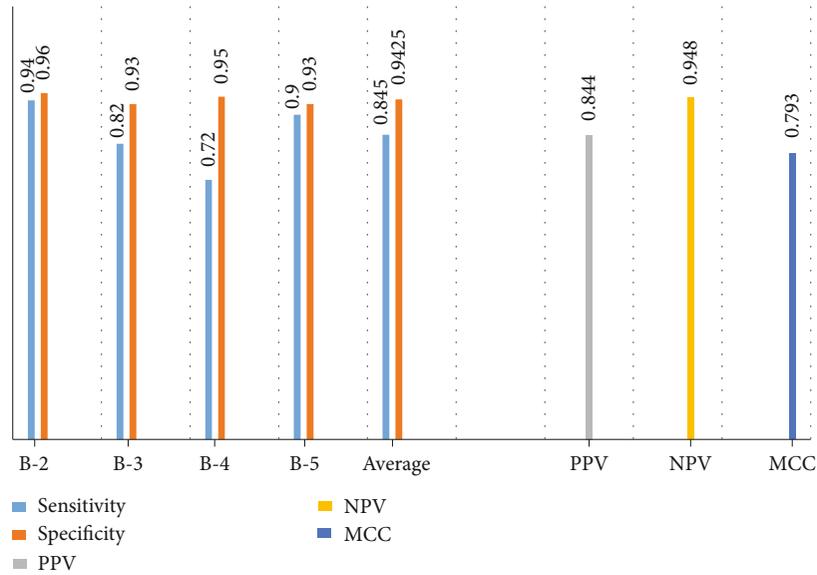

Figure 8: Evaluation metrics computed from the best classification result.

Table 3: Comparison with existing state-of-the art counterparts.

| Method | Dataset | A_Sn (%) | A_Sp (%) | PPV (%) | NPV (%) | MCC (%) | Accuracy (%) | Year |
|---|---|---|---|---|---|---|---|---|
| [27] | 46 DDSM | 76 | 84 | — | — | — | 76.67 | 2015 |
| [18] | 480 DDSM | 83 | 94 | 83.9 | 94.6 | 78.4 | 83.85 | 2016 |
| [28] | 410 INbreast | — | — | — | — | — | 83.4 | 2018 |
| Ours | 500 DDSM | 84.5 | 94.25 | 84.4 | 94.8 | 79.3 | 84.5 | 2020 |

A_Sn: average sensitivity; A_Sp: average specificity.

in the diagonal of Table 2 and are marked in bold. We can see that from a total of 200 testing samples, our approach successes to classify 169 samples and fails to recognize 31 samples. Thus, the overall accuracy achieved is 84.5%, and the individual classification rates obtained for each of the four BI-RADS categories B-2, B-3, B4, and B-5 are 94%, 82%, 72%, and 90%, respectively. As a result, the benign category (B-2) obtained the highest classification accuracy followed by the highly suggestive of malignancy (B-5). The main reason behind it is that they often involve more representative BI-RADS features that have capabilities to discriminate benign and malignant categories (mostly interpreted as B-2 and B-5 in BI-RADS).

Moreover, it is clear that the BPN classifier showed acceptable capabilities to classify the probably benign category (B-3) with 82% classification accuracy. However, it faced difficulties to classify the suspicious finding category (B-4) as it has received the lowest classification accuracy with 72%. This limitation could be justified by the huge overlap existing between (B-3/B-4) categories and between (B4/B-5) categories as well. As shown in Figure 4, we can observe from the proposed semiautomatic segmentation results the big similarities between the B-3 and B-4 segmented ROIs in terms of shape and margin, with a slight difference in the texture, which makes the discrimination between them much more complex and confusing.

Furthermore, Figure 8 shows the sensitivity and the specificity for each BI-RADS category and their average, as well as the overall obtained values of PPV, NPV, and MCC. As can be seen, we achieved PPV, NPV, and MCC of 84.4%, 94.8%, and 79.3%, respectively. Meanwhile, our proposed approach yielded an average sensitivity and specificity of 84.5% and 94.25%, respectively.

4.4. Comparison with State-of-the-Art Counterparts. Our approach is compared with the works of Miranda and Felipe [27], Chokri and Farida [18], and Domingues et al. [28]. They are few publications that have reported the BI-RADS classification of breast masses using mammographic images. The evaluation comparisons are shown in Table 3. As can be seen, our proposed CAD system exhibits superiority in terms of overall classification accuracy and provides better results in terms of average sensitivity, average specificity, PPV, NPV, and MCC.

4.5. Effectiveness of the Modified Genetic Feature Selection Method. In this section, we demonstrate the effectiveness of our modified genetic feature selection method. The training and testing accuracies versus the best feature subsets obtained by using this method are depicted in Figure 9. Note that the x-axis represents the best feature subset for each possible number of features that provides the highest classification accuracy. The graphs show that initially with the increasing number of the included features, the training and testing accuracies of the BPN classifier increase. However, they jointly decrease when more features are added. This is justified by the fact that some included features are not significant and redundant and may lead to the performance's deterioration.

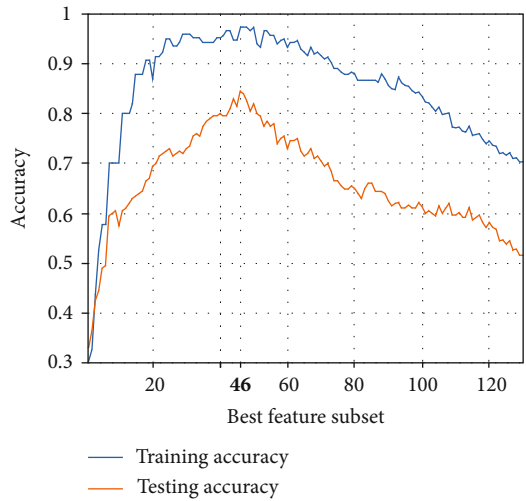

Figure 9: Training and testing accuracies versus the best feature subset.

More specifically, the best feature subset that provides the highest classification accuracy for distinguishing the four BI-RADS categories is found with 46 features. Moreover, no single BI-RADS feature can be sufficient for the mammogram mass classification, as the best single feature (*difference entropy* "*standard deviation*") yields the training accuracy of 30.33% with a maximum testing accuracy of 33% which is not practically acceptable. We can also notice that the classification accuracy in the testing data has been significantly increased from 51.5% (with training accuracy of 70.33%) for the full original features to 84.5% (with training accuracy of 97.33%) for the 46 best selected features, which can validate the effectiveness of our proposed method to enhance the classification performance of the BI-RADS-based mass classification.

In addition, to have a glance at the 46 handcrafted BI-RADS features selected by our designed genetic feature selection, which provided the highest classification accuracy, we summarize them in Table 4.

From Table 4, we can derive some important notes:

(1) Different numbers of features are selected from each BI-RADS category (i.e., 5 shape features, 9 margin features, and 30 textural features). This explains, in particular, the importance and contribution of each type of features to the performance of the BI-RADS mass classification. In addition, the 46 selected features are complementary to each other and cover all BI-RADS categories

(2) The selection of the additional features (mass size and patient's age) reinforces what is mentioned in the BI-RADS atlas [7] concerning the crucial significance and importance of these two features for breast cancer classification into BI-RADS categories

(3) The best single feature (*difference entropy* "*standard deviation*") is not present on the best 46 feature subsets, which explains its irrelevance when combined with other features

(4) Using our modified genetic feature selection, more textural features compared with shape and margin features have been selected, which may explain the wide varieties of textures in mammogram masses that require more textural features to recognize them

Based on these results, we believe that our modified genetic feature selection method selects more significant BI-RADS features for classifying border categories (B-2/B-5) and less significant BI-RADS features for classifying in-between categories (B-3/B-4).

Another important result from our proposed approach is the encouraging classification rate (82%) obtained for the probably benign category (B-3) which has been the focus of public debate in screening mammography and is considered as the most difficult category to use in practice [8, 47]. In order to improve the capacities of the BPN classifier for distinguishing the in-between categories, we recommend an extraction of more meaningful features (especially for shape and margin) by using a fusion of multiple feature extraction techniques and further applying our proposed feature selection method to figure out the most representative features for classification.

## 5. Conclusions

In this paper, we have proposed a new and effective CAD system to classify mammogram masses into four BI-RADS categories (B2, B-3, B-4, and B-5), which can support the radiologists' diagnosis. The mammogram images are first enhanced by using histogram equalization. To ensure good segmentation results, we semiautomatically segment the ROIs based on the region growing technique. A set of 130 handcrafted BI-RADS features has been then extracted mainly from the mass shape, margin, and density. Thereafter, we proposed a modified genetic feature selection method, where GA is designed with problem-specific operators and applied successively over all the possible number of features to search for the optimal feature subsets. Then, the global optimal subset is automatically obtained. The overall accuracy has been significantly improved as we achieved 84.5% from the best feature subset including 46 handcrafted features. More specifically, the benign category obtained the highest classification accuracy with 94%, followed by the highly suggestive of malignancy with 90%, while the suspicious finding received the lowest accuracy with 72%. We also obtained the PPV, NPV, and MCC of 84.4%, 94.8%, and 79.3%, respectively, which outperform the state-of-the-art counterparts. Based on the significance of the selected BI-RADS features, hopefully, the proposed system can support radiologist decisions for determining proper patient

TABLE 4: Summary of the 46 selected features as the best feature subset using our modified genetic feature selection method.

| BI-RADS category | Feature name | Feature significance |
|---|---|---|
| Shape | Irregularity<br>Difference area: convex hull area minus mass area<br>Variance variation<br>Kurtosis variation<br>Entropy variation | (i) Irregularity: this feature is extracted based on active contour method (aka snakes) [39]. When the edge path of a mammographic mass changes its direction, a point $p_i$ should be highlighted. The irregularity is therefore equal to the number of points $p_i$ along the margin.<br>(ii) The remaining features (difference area: convex hull area minus mass area, variance variation, kurtosis variation, and entropy variation) are proposed in sonography by [40]. The convex hull is the smallest convex polygon that contains the mass contour and the mass region. The convex hull area is defined as the actual number of pixels in the convex hull of the mass, and the mass area is defined as the actual number of pixels in the mass region. To compute the other features, a function called "variation" which is the projection of the distance between the farthest pixels of a mass region at all angles is used, where variance, kurtosis, and entropy are statistical values of it. |
| Margin | Kurtosis "variance" kurtosis "Kurtosis"<br>Entropy "minimum"<br>Entropy "average"<br>Entropy "variance"<br>Entropy "kurtosis"<br>Index of the maximum probability "minimum"<br>Index of the maximum probability "maximum"<br>Index of the maximum probability "variance" | (i) These features are quantified from a set of waveforms by wavelet analysis used in [41].<br>(ii) Margin kurtosis: it measures how much peaked is a probability distribution, and for well-defined margins, it gets higher values. In general, well-defined margins have an abrupt transition along a waveform.<br>(iii) Margin entropy is defined as a state of disorder or decline into disorder of edges waveforms. It is also expected that well-defined margins have lower entropy.<br>(iv) Index of the maximum probability is another measure which is shifted to be zero on the margin, positive for outside, and negative for inside of the margin. This index is used to capture variations of most probable edge places among the margin.<br>(v) A waveform with length $l = 64$ is placed sequentially while traversing the margin, and accordingly, an edge probability vector (EP) is computed for each waveform $i$. Then, the features "kurtosis," "entropy," and "the index of the maximum probability" are calculated from EP for each waveform $i$. However, as we set the number of waveforms to 32 in our experiments, statistical functions such as variance, kurtosis, minimum, maximum, and average are used to reduce the number of margin features. |
| Density | Angular second moment "skewness"<br>Contrast "minimum"<br>Contrast "average"<br>Contrast "variance"<br>Contrast "standard deviation"<br>Contrast "skewness"<br>Correlation "variance"<br>Correlation "standard deviation"<br>Correlation "skewness"<br>Variance (sum of squares) "average"<br>Variance (sum of squares) "variance"<br>Variance (sum of squares) "skewness"<br>Inverse difference moment "variance"<br>Sum average "maximum"<br>Sum average "standard deviation"<br>Sum variance "maximum"<br>Sum entropy "maximum"<br>Sum entropy "kurtosis"<br>Entropy "minimum"<br>Difference variance "variance"<br>Difference entropy "maximum"<br>Difference entropy "average"<br>Difference entropy "variance"<br>Information measure of correlation 1 "minimum" | Is a measure of the asymmetry of the "angular second moment feature" distribution about its mean.<br>Is the smallest observation of "contrast features."<br>Is the statistical value that describes the center of a set of "contrast features."<br>It measures how far a set of "contrast features" are spread out from their average value.<br>Is a measure of the amount of variation or dispersion of a set of "contrast features."<br>Is a measure of the asymmetry of the "contrast features" distribution about its mean.<br>It measures how far a set of "correlation features" is spread out from their average value.<br>Is a measure of the amount of variation or dispersion of a set of "correlation features."<br>Is a measure of the asymmetry of the "correlation features" distribution about its mean.<br>Is the statistic value which describes the center of a set of "sum of square features."<br>It measures how far a set of "sum of square features" is spread out from their average value.<br>Is a measure of the asymmetry of the "sum of square features" distribution about its mean.<br>It measures how far a set of "inverse difference moment features" is spread out from their average value.<br>Is the greatest observation of "sum average features."<br>Is a measure of the amount of variation or dispersion of a set of "sum average features."<br>Is the greatest observation of "sum variance features."<br>Is the greatest observation of "sum entropy features."<br>Is a measure of whether the "sum entropy features" are heavy-tailed or light-tailed relative to a normal distribution.<br>Is the smallest observation of "entropy features."<br>It measures how far a set of "difference variance features" is spread out from their average value.<br>Is the greatest observation of "difference entropy features."<br>Is the statistical value that describes the center of a set of "difference entropy features."<br>It measures how far a set of "difference entropy features" is spread out from their average value.<br>Is the smallest observation of the "information measure of correlation 1 features." |

TABLE 4: Continued.

| BI-RADS category | Feature name | Feature significance |
|---|---|---|
| | Information measure of correlation 1 "standard deviation" | Is a measure of the amount of variation or dispersion of a set of "information measure of correlation 1 features." |
| | Information measure of correlation 1 "skewness" | Is a measure of the asymmetry of the "information measure of correlation 1 feature" distribution about its mean. |
| | Information measure of correlation 2 "minimum" | Is the smallest observation of the "information measure of correlation 2 features." |
| | Information measure of correlation 2 "maximum" | Is the greatest observation of the "information measure of correlation 2 features." |
| | Maximal correlation coefficient "standard deviation" | Is a measure of the amount of variation or dispersion of a set of "maximal correlation coefficient features." |
| | Maximal correlation coefficient "kurtosis" | Is a measure of the combined weight of the "maximal correlation coefficient feature" distribution's tails relative to the center of the distribution. |
| Additional features | Mass size | (i) The mass size is represented by the zone included inside the contour and is computed in $mm^2$. |
| | Patient age | (ii) The patient's age is also used and is the unique human feature used here. |

management based on the assigned BI-RADS categories. We also anticipate that our approach will not only help to enhance the current clinical mammogram assessment of breast lesions BI-RADS categorization but also can be further generalized to other medical classification applications.

In our future work, a deep architecture using CNN to learn BI-RADS features automatically from raw data images will be explored and a combination of our designed feature selection method with automatic extracted features through deep CNN will be investigated to enhance the BI-RADS-based classification performance.

## Data Availability

The mammogram image data (DDSM) used to support the findings of this study is public and freely available on web. The link for accessing this data is provided within the article, in Section 4.1.

## Conflicts of Interest

The authors declare that they have no conflicts of interest.

## Acknowledgments

This work was supported in part by the Beijing Municipal Science and Technology Project (grant number Z181100001918002) and the Fundamental Research Funds for the Central Universities and the PUMC Youth Fund (Grant numbers 2017320010 and 3332019009).